\documentclass[a4paper]{jpconf}

\usepackage{verbatim}
\usepackage{graphicx}
\usepackage{dcolumn}
\usepackage{bm}
\usepackage{color}
\usepackage{url}
\usepackage{amsmath}
\usepackage{morefloats}

\usepackage{color}

\newcommand{\be}{\begin{equation}}
\newcommand{\ee}{\end{equation}}
\newcommand{\ba}{\begin{eqnarray}}
\newcommand{\ea}{\end{eqnarray}}

\newcommand{\Mc}{{\cal M}}
\newcommand{\Ms}{M_{\odot}}

\newcommand{\hyp}{\mathcal{H}}
\newcommand{\info}{\mathrm{I}}

\def\ltsima{$\; \buildrel < \over \sim \;$}
\def\simlt{\lower.5ex\hbox{\ltsima}}
\def\gtsima{$\; \buildrel > \over \sim \;$}
\def\simgt{\lower.5ex\hbox{\gtsima}}
\begin{document}

\title[title]{Towards a generic test of the strong field dynamics of general relativity using compact binary coalescence: Further investigations}

\author{T.G.F.~Li$^{1}$, W.~Del Pozzo$^{1}$, S.~Vitale$^{1}$, C.~Van Den Broeck$^{1}$, M.~Agathos$^{1}$, J.~Veitch$^{1,2}$, K.~Grover$^{3}$, T.~Sidery$^{3}$, R.~Sturani$^{4,5}$, A.~Vecchio$^{3}$}
\address{$^1$Nikhef -- National Institute for Subatomic Physics, Science Park 105, 1098 XG Amsterdam, The Netherlands \\
$^2$School of Physics and Astronomy, Cardiff University, Queen's Buildings, The Parade, Cardiff CF24 3AA, United Kingdom\\
$^3$School of Physics and Astronomy, University of Birmingham, Edgbaston, Birmingham B15 2TT, United Kingdom\\
$^4$Dipartmento di Scienze di Base e Fondamenti, Universit\`a di Urbino, I-61029 Urbino, Italy\\
$^5$INFN, Sezione di Firenze, I-50019 Sesto Fiorentino, Italy}
\date{\today}

\begin{abstract}

In this paper we elaborate on earlier work by the same authors in which a novel Bayesian inference framework for testing the strong-field dynamics of General Relativity using coalescing compact binaries was proposed. Unlike methods that were used previously, our technique addresses the question whether \emph{one or more} `testing coefficients' (\emph{e.g.} in the phase) parameterizing deviations from GR are non-zero, rather than all of them differing from zero at the same time. The framework is well-adapted to a scenario where most sources have low signal-to-noise ratio, and information from multiple sources as seen in multiple detectors can readily be combined. In our previous work, we conjectured that this framework can detect \emph{generic} deviations from GR that can in principle not be accomodated by our model waveforms, on condition that the change in phase near frequencies where the detectors are the most sensitive is comparable to that induced by simple shifts in the lower-order phase coefficients of more than a few percent ($\sim 5$ radians at 150 Hz). To further support this claim, we perform additional numerical experiments in Gaussian and stationary noise according to the expected Advanced LIGO/Virgo noise curves, and coherently injecting signals into the network whose phasing differs structurally from the predictions of GR, but with the magnitude of the deviation still being small. We find that even then, a violation of GR can be established with good confidence.
\end{abstract}



\section{Introduction}
\label{sec:introduction} 
The theory of General Relativity (GR) is currently the most commonly accepted theory describing space-time and gravitation. The theory has been accurately tested within the weak-field, stationary regime (\emph{e.g.}~through Solar System tests \cite{Misner:1973kx}), and no deviations from GR have been conclusively found. However, by the very nature of GR, gravitation is a dynamical phenomenon, a key aspect being the prediction of gravitational waves (GW) \cite{Will:2006zr, Sathyaprakash:2009ly}. The first clue towards this dynamical aspect came with the discovery of the Hulse-Taylor binary pulsar \cite{Hulse:1975ve}. Its orbital motion is in agreement with the assumption that GWs carry away energy and angular momentum as predicted by GR. This test, as well as similar test performed on other binary pulsars that were subsequently discovered \cite{Burgay:2003qf,Kramer:2006bh}, however still involve indirect observations of GWs. Furthermore, even for the most relativistic binary pulsar that is currently known (PSR J0737-3039), one has $G M/(c^2 R) \simeq 4.4 \times 10^{-6}$ (where $M$ is the total mass, and $R$ is the orbital separation) and a typical orbital speed of $v/c \simeq 2 \times 10^{-3}$. The observed binary pulsars are still very far from merger. However,
given access to a sufficiently large volume of space, one should find 
compact binaries in the final stages of inspiral. At the nominal last stable orbit, these will have a separation of $R = 6 G M/c^2$, $G M/(c^2 R) = 1/6$ and $v/c = 1/\sqrt{6}$. The process of inspiral has been modelled in detail in the context of the so-called post-Newtonian (PN) approximation (\emph{cf}.~\cite{Blanchet:2006fk} and references therein). The direct detection of the gravitational wave signals from such binaries would enable us to probe the genuinely strong-field, dissipative dynamics of GR. 

Compact binary systems that are close to merger are in fact among the primary targets for kilometer-sized interferometric detectors. These include the Virgo detector in Italy \cite{Acernese:2008oq, Accadia:2011kl}, the two LIGO detectors in the United States \cite{Abbott:2009dq} and GEO600 in Germany \cite{Grote:2010hc}. Both Virgo and LIGO are in the process of being upgraded to Advanced Virgo and Advanced LIGO, which are expected to be completed around 2015 \cite{Harry:2010ij, LIGO-Scientific-Collaboration:fu,Virgo-Collaboration:2009ys,Virgo_collaboration:fu}. Furthermore, an interferometer named KAGRA (formerly know as LCGT) in Japan is in a planning stage \cite{Kuroda:2010mi}, and a detector in India \cite{Sathyaprakash:2011uq} is also being considered. With the current estimates for the advanced detectors, the rate of detection of inspiralling compact binaries is expected be around a few tens per year \cite{Abadie:2010zr}.

Quite a number of alternative theories to GR have been discussed in the literature, and the accuracy with which some of these could be probed with gravitational waves has been studied within the Fisher matrix formalism; for scalar-tensor theories, see \cite{Will:1994gb, Damour:1998mb, Scharre:2002cq, Will:2004kh, Berti:2005rq, Berti:2005fc}, for a varying Newton constant \cite{Yunes:2010ss}, for modified dispersion relation theories (commonly referred to as `massive gravity') \cite{Berti:2005rq, Berti:2005fc, Will:1998kx, Arun:2009vn, Stavridis:2009ys, Keppel:2010zr, Berti:2011ve}, for violations of the No Hair Theorem \cite{Berti:2006qf, Hughes:2006bh, Barack:2007dq, Gair:2008cr, Kamaretsos:2011nx}, for violations of Cosmic Censorship \cite{Kamaretsos:2011nx, Van-Den-Broeck:2007zh}, and for parity violating theories \cite{Alexander:2008kl, Yunes:2009tg, Sopuerta:2009hc, Yunes:2010ij}. Practical Bayesian methods for performing tests of GR on actual gravitational wave data when they become available include the work by Del Pozzo \emph{et al.} \cite{Del-Pozzo:2011qf} in the context of massive gravitons, that of Cornish \emph{et al.} \cite{Cornish:2011fk}, which employed the so-called parameterized post-Einsteinian (ppE) waveform family \cite{Yunes:2009fu,Yunes:2010zt}, and that of Gossan \emph{et al.} \cite{Gossan:2011} which focussed on the ringdown signal. 

A proposal by Arun \emph{et al.} \cite{Arun:2006ys,Arun:2006dz,Mishra:2010bs} is to measure various quantities within the phase and to check their consistency with the predictions of GR. This could lead to a very generic test, in that one would not be looking for particular (classes of) alternative theories. However, so far its viability was only explored through Fisher matrix studies.

Inspired by the method of Arun \emph{et al.}, the authors of the present paper developed a new Bayesian framework, with the following features \cite{Li:2011fk}:
\begin{itemize}
\item Contrary to previous Bayesian treatments such as \cite{Cornish:2011fk,Gossan:2011}, it addresses the question ``Do \emph{one or more} testing parameters characterizing deviations from GR differ from zero?'' as opposed to ``Do \emph{all} of them differ from zero?'' In practice this comes down to testing a number of auxiliary hypotheses, in each of which only a subset of the set of testing parameters is allowed to be non-zero;
\item Precisely because in most of the auxiliary hypotheses a smaller set of testing parameters is used, this method will be more suited to a scenario where most sources have low signal-to-noise ratio, as we expect to be the case with Advanced LIGO/Virgo;
\item As with most Bayesian methods, information from multiple sources can easily be combined;
\item The framework is not tied to any particular waveform family or even any particular part of the coalescence process. However, in \cite{Li:2011fk} we focussed on the inspiral part and chose the testing parameters to be shifts in the lower-order inspiral phase coefficients, as we will also do here.
\end{itemize}

Besides establishing a theoretical framework, \cite{Li:2011fk} also showed results for a few simple example deviations from GR. In particular, it was illustrated how the method can be sensitive to deviations which in principle cannot be accomodated by the model waveforms. In fact, it is reasonable to assume that the technique will be able to pick up \emph{generic} deviations from GR, on condition that their effect on the phasing is of the same magnitude as that of a simple shift in one or more of the low-order phasing coefficients of the standard post-Newtonian waveform. More precisely, as long as the change in the phase at frequencies where the detectors are the most sensitive is comparable to the effect of a shift of a few percent at $(v/c)^3$ beyond leading order (corresponding to $\sim 5$ radians at $f = 150$ Hz), we expect the deviation from GR to be detectable by our method. In this paper, we will show some striking examples to provide further support for this claim.

Subsequent sections of this paper are structured as follows. In section \ref{sec:method} we recall the theory and the implementation of the method introduced in \cite{Li:2011fk}. Section \ref{sec:results} shows results from simulations done with some specific examples of modifications to the waveform phase. A discussion and conclusions are presented in section \ref{sec:conclusion_discussion}.


\section{Method}
\label{sec:method}

\subsection{Bayesian Inference}
\label{subsec:inference}

At the heart of the method we proposed in \cite{Li:2011fk} lies the question ``to what degree do we believe GR is the correct theory describing the detected signals?'' This question is best answered within the framework of Bayesian model selection \cite{Jaynes:2003qa}. The cornerstone of Bayesian analysis is the comparison between the probabilities of two hypotheses given the available data. This is quantified by the \emph{odds ratio}
\begin{equation}
	O^i_j = \frac{P({\cal H}_i|d, \info)}{P({\cal H}_j|d, \info)},
\end{equation}
where ${\cal H}_i, {\cal H}_j$ are the hypotheses of the models to be compared, $d$ represents the data and $\info$ is the relevant background information. Using Bayes' theorem, we can then write this odds ratio as
\begin{equation}
	O^i_j = \frac{P(d | {\cal H}_i, \info)}{P(d | {\cal H}_j, \info)} \frac{P({\cal H}_i| \info)}{P({\cal H}_j|\info)} = B^i_j \frac{P({\cal H}_i| \info)}{P({\cal H}_j|\info)}.
\label{OddsRatio}
\end{equation}
The odds ratio is thus the product of two ingredients. The first factor is the ratio of the so-called \textit{evidences}, $B^i_j = \frac{P(d | {\cal H}_i, \info)}{P(d | {\cal H}_j, \info)}$, which is also known as the \textit{Bayes factor}. The evidence (also called the marginal likelihood) for \emph{e.g.} the hypothesis $\mathcal{H}_i$ is given by
\begin{equation}
	\label{eq:evidence}
	P(d | {\cal H}_i, \info) = \int d\vec{\theta}\, p(\vec{\theta} | {\cal H}_i, \info)\, p(d | \vec{\theta}, {\cal H}_i, \info)\,
\end{equation}
where $\vec{\theta}$ are the parameters associated with the hypothesis ${\cal H}_i$, and $p(\vec{\theta} | {\cal H}_i, \info)$ is the prior probability distribution of the parameters.

The second factor in Eq.~(\ref{OddsRatio}) is the ratio of the prior probabilities, $\frac{P({\cal H}_i| \info)}{P({\cal H}_j|\info)}$, and is often referred to as \textit{prior odds}. It should be noted that the prior probability distribution is uniquely determined by the prior information.

The assignment of the prior probability will be further explained in subsection \ref{subsec:odds_ratio}. Details on the calculation of the Bayes factor can be found in subsection \ref{subsec:implementation}.

\subsection{Waveform model}
\label{subsec:waveform_model}

Before moving on to define the odds ratio for the problem at hand, let us explain the model waveforms used in this paper. In the inspiral regime of compact binary coalescing systems, the waveforms are accurately described by the post-Newtonian approximation. This approximation describes important quantities such as the energy and the flux as expansions in powers of $v/c$, where $v$ is the characteristic velocity of the binary system.


To illustrate our method we will use the analytic, frequency domain TaylorF2 waveform model \cite{Thorne:1987ff, Sathyaprakash:1991lh}, which is implemented in the LIGO Algorithms Library in the following way \cite{LAL}:
\begin{equation}
	h(f) = \frac{1}{D} \frac{\mathcal{A}(\theta, \phi, \iota, \psi, \mathcal{M},\eta)}{\sqrt{\dot{F}(\mathcal{M},\eta;f)}} f^{2/3}\,e^{i\Psi(\mathcal{M},\eta;f)},
	\label{TaylorF2}
\end{equation}
where $D$ is the distance, $(\theta,\phi)$ the sky position in the detector frame, and $(\iota,\psi)$ the orientation of the orbital plane with respect to the direction to the line of sight. $\mathcal{M}$ is the so-called chirp mass, and $\eta$ is the symmetric mass ratio; in terms of the component masses $(m_1, m_2)$ one has $\eta = m_1 m_2/(m_1 + m_2)^2$ and $\mathcal{M} = (m_1 + m_2)\,\eta^{3/5}$. The 
phase $\Psi(\mathcal{M}, \eta; f)$ takes the form
\begin{equation}
	\label{psiidef}
	\Psi(\mathcal{M},\eta;f) = 2\pi f t_c - \phi_c - \pi/4 + \sum_{i=0}^7 \left[\psi_i + \psi_{i}^{(l)} \ln f \right]\,f^{(i-5)/3},
\end{equation}
with $t_c$ and $\phi_c$ the time and phase at coalescence, respectively. Central to the method are the phase coefficients $\psi_i$ and $\psi_{i}^{(l)}$. These either have a functional dependence on $(\mathcal{M},\eta)$ as predicted by GR (\textit{cf.} \cite{Mishra:2010bs} for the explicit expressions), or are allowed to deviate from the value predicted by GR.
In Eq.~(\ref{TaylorF2}), the `frequency sweep' 
$\dot{F}(\mathcal{M},\eta;f)$ is itself an expansion in powers of the frequency $f$ with mass-dependent coefficients.  
Note that $\dot F$ is related to the phase $\Psi$ and we could in principle
allow it to deviate from the GR prediction. However, for stellar mass binaries and
with advanced detectors, we do not expect to be particularly sensitive to sub-dominant
contributions to the amplitude \cite{Van-Den-Broeck:2007zh}, so we will keep the function $\dot{F}$ fixed 
to its GR expression.


We note that in the case of binary neutron stars, which are the sources we will in fact focus on, TaylorF2 waveforms are likely to already suffice for a first test of GR. Indeed, in the relevant mass range, TaylorF2 has a match and fitting factor close to unity with Effective One-Body waveforms modified for optimal agreement with numerical simulations \cite{Buonanno:2009mi}. Spins are unlikely to be very important in this case. One might worry about finite size effects, but as shown in \cite{Hinderer:2010}, even for the most extreme neutron star equations of state and for sources as close as 100 Mpc, advanced detectors will not be sensitive to these at frequencies below $\sim 400$ Hz; hence one could cut off the recovery waveforms at 400 Hz, in which case the loss in SNR would be less than a percent. However, if one also wanted to test GR using systems composed of a neutron star and a black hole, or two black holes, then dynamical spins \cite{Blanchet:1995il}, sub-dominant signal harmonics \cite{Blanchet:2008et,Van-Den-Broeck:2007zh} and merger/ringdown \cite{Pan:2011yb} would become important, and in that case more sophisticated waveform models will be called for. The latter is currently the subject of investigation.

\subsection{Defining the odds ratio}
\label{subsec:odds_ratio}

Focussing our attention on deviations from GR in the phase of the measured 
waveform, \emph{i.e.} keeping the amplitude fixed to its GR-predicted value, 
we consider within the Bayesian model selection framework the following two  
hypotheses:
\begin{itemize}
\item $\hyp_{\rm GR}$: The waveform has a phase with the functional dependence on 
$(\mathcal{M},\eta)$ as predicted by GR;
\item $\hyp_{\rm modGR}$: One or more of the phase coefficients do not have the functional dependence on $(\mathcal{M},\eta)$ as predicted by GR.
\end{itemize}
The GR hypothesis, $\hyp_{\rm GR}$, is the hypothesis that our GR waveform model (TaylorF2 in this case) correctly describes the signal originating from the inspiral of two compact objects. Ideally, $\hyp_{\rm modGR}$ would simply have been the negation of $\hyp_{\rm GR}$. However, \emph{a priori}, deviations from GR can occur an infinite number of ways. What we will argue is that for the core question that we want to address, \textit{i.e.} whether or not the observed phase deviates from GR, it will be sufficient to allow for a \emph{limited} set of possible deviations in the recovery waveforms. 
For TaylorF2, we take the set of deviations only to be within the known phase coefficients $\{\psi_{0}, \psi_{1}, \psi_{2}, \ldots, \psi_N \}$.

To date, the TaylorF2 phase has ten known phase coefficients ($\psi_0, \ldots, \psi_7$, and two additional coefficients $\psi^{(l)}_5$ and $\psi^{(l)}_6$ associated with logarithmic contributions). Here we will not use $\psi_0$ as a variable coefficient; even so, if one were to consider all the subsets of the set of remaining  coefficients, one would have to take into account $2^9 - 1 = 511$ ways in which a deviation can occur. Apart from this being computationally demanding, we do not expect to be sensitive to the highest-order coefficients; hence it makes sense to limit oneself to all the subsets of
\begin{equation}
	\{\psi_{1}, \psi_{2}, \ldots, \psi_{N_T} \},
\end{equation}
where $N_T$ is the number of phase coefficients one chooses to consider. We thus allow one or more of the coefficients $\{\psi_{1}, \psi_{2} \ldots, \psi_{N_T}\}$ to vary freely, instead of following the functional dependence on $(\Mc, \eta)$ as predicted by GR. The choice of $N_T$ will be in part be influenced by the required generality of the test, measurability of phase coefficients, and computational limitations. 

Finally, we quantify our belief in whether one or more phase coefficients deviate from GR by means of auxiliary hypotheses $H_{i_1 i_2 \ldots i_k}$, which are defined as follows:
\begin{quote}
$H_{i_1 i_2 \ldots i_k}$ is the hypothesis that the phasing coefficients $\psi_{i_1}, \ldots, \psi_{i_k}$ do not have the functional dependence on $(\Mc,\eta)$ as predicted by General Relativity, but all other coefficients $\psi_j$, $j \notin \{i_1, i_2, \ldots, i_k\}$ \emph{do} have the dependence as in GR.
\end{quote}
It is important to note that by definition of the hypotheses $H_{i_1 i_2 \ldots i_k}$, they are mutually, logically disjoint, \textit{i.e.}, $H_{i_1 i_2 \ldots i_k} \wedge H_{j_1 j_2 \ldots j_l}$ is always false for $\{i_1, i_2, \ldots, i_k\} \neq \{j_1, j_2, \ldots, j_l\}$.

For a signal to be inconsistent with GR, we require that one or more phase coefficients deviate from GR. In terms of hypotheses, we are thus interested in the logical `or' of the sub-hypotheses, $H_{i_1 i_2 \ldots i_k}$, defined above. With this in hand, we can now define $\hyp_{\rm modGR}$ to be:
\begin{equation}
	\hyp_{\rm modGR} = \bigvee_{i_1 < i_2 < \ldots < i_k; k \leq N_T} H_{i_1 i_2 \ldots i_k}.
	\label{eq:logicalOR}
\end{equation}
The odds ratio for $N_T$ coefficients is given by:
\begin{equation}
	{}^{(N_T)}O^{\rm modGR}_{\rm GR} = \frac{P(\hyp_{\rm modGR}|d,\info)}{P(\hyp_{\rm GR}|d,\info)} = \frac{P(\bigvee_{i_1 < i_2 < \ldots < i_k; k \leq N_T} H_{i_1 i_2 \ldots i_k}|d,\info)}{P(\hyp_{\rm GR}|d,\info)},
\end{equation}
Using the fact that the auxiliary hypotheses are mutually, logically disjoint, one can write
\begin{equation}
	P(\bigvee_{i_1 < i_2 < \ldots < i_k; k \leq N_T} H_{i_1 i_2 \ldots i_k}|d,\info) = \sum_{k=1}^{N_T} \sum_{i_1 < i_2 < \ldots < i_k} P(H_{i_1 i_2 \ldots i_k}|d,\info).
\end{equation}
Applying Bayes' theorem, one finds
\begin{equation}
	{}^{(N_T)}O^{\rm modGR}_{\rm GR} = \sum_{k=1}^{N_T} \sum_{i_1 < i_2 < \ldots < i_k} \frac{P(H_{i_1 i_2 \ldots i_k}|\info)}{P(\hyp_{\rm GR}|\info)} B^{i_1 i_2 \ldots i_k}_{\rm GR},
\end{equation}
where 
\begin{equation}
	B^{i_1 i_2 \ldots i_k}_{\rm GR} = \frac{P(d|H_{i_1 i_2 \ldots i_k},\info)}{P(d|\hyp_{\rm GR},\info)}.
\end{equation}
At this point, one has to set the values for the relative prior probabilities, $\frac{P(H_{i_1 i_2 \ldots i_k}|\info)}{P(\hyp_{\rm GR}|\info)}$. When one is devoid of prior information as to which of the test coefficients are inconsistent with GR, one can choose to invoke total ignorance and assign to  each an equal weight, \textit{i.e.}
\begin{equation}
	\frac{P(H_{i_1 i_2 \ldots i_k}|\info)}{P(\hyp_{\rm GR}|\info)} = \frac{P(H_{j_1 j_2 \ldots j_l}|\info)}{P(\hyp_{\rm GR}|\info)}\,\,\,\,\,\,\,\mbox{for any $k, l \leq N_T$}.
	\label{eq:generalprioroddsratios}
\end{equation}
Despite the choice of total ignorance, however, one more quantity needs to be set. The overall relative prior, $\frac{P(\hyp_{\rm modGR}|\info)}{P(\hyp_{\rm GR}|\info)}$, describes the prior belief in whether GR is the correct theory or not. The choice of this quantity is left to the reader. For convenience, however, we write
\begin{equation}
	\frac{P(\hyp_{\rm modGR}|\info)}{P(\hyp_{\rm GR}|\info)} = \alpha.
	\label{eq:totalignorance}
\end{equation}
As will become apparent below, $\alpha$ will end up being just an overall scaling of the odds ratio. Later on, for the purposes of showing results, we will set $\alpha = 1$.

The equality (\ref{eq:totalignorance}), together with (\ref{eq:generalprioroddsratios}) and the logical disjointness of the $2^{N_T} - 1$ hypotheses $H_{i_1 \ldots i_k}$ implies
\begin{equation}
	\frac{P(H_{i_1 i_2 \ldots i_k}|\info)}{P(\hyp_{\rm GR}|\info)} = \frac{\alpha}{2^{N_T} - 1}.
	\label{eq:component-odds}
\end{equation}
In terms of the $H_{i_1 i_2 \ldots i_k}$, the odds ratio can then be written as
\begin{equation}
	{}^{(N_T)}O^{\rm modGR}_{\rm GR} = \frac{\alpha}{2^{N_T}-1}\,\sum_{k=1}^{N_T} \sum_{i_1 < i_2 < \ldots < i_k} B^{i_1 i_2 \ldots i_k}_{\rm GR}.
	\label{eq:oddsindividual}
\end{equation}
Up to an overall prefactor, the odds ratio is thus a straightforward average of the Bayes factors from the individual sub-hypotheses, $H_{i_1 i_2 \ldots i_k}$.

\subsection{Multiple sources}
\label{subsec:multiple_sources}

Although the detection rate for compact binary coalescences is still rather uncertain, we expect advanced instruments to detect several events per year~\cite{Abadie:2010zr}. It is therefore important to take advantage of multiple detections to provide tighter constraints on the validity of GR.

The extension of the odds ratio to include observations from several independent sources can be found in \cite{Del-Pozzo:2011qf, Li:2011fk}. Here we simply state the result, referring the interested reader to these papers for details. If one assumes $\mathcal{N}$ independent measurements and the events are labelled by $A$, one can write the odds ratio as

\begin{equation}
	{}^{(N_T)}\mathcal{O}^{\rm modGR}_{\rm GR} =  
	\frac{\alpha}{2^{N_T} -1}\,\sum_{k=1}^{N_T} \sum_{i_1 < i_2 < \ldots < i_k} \prod_{A=1}^\mathcal{N} {}^{(A)}B^{i_1 i_2 \ldots i_k}_{\rm GR},
	\label{eq:oddscombined}
\end{equation}
where
\begin{equation}
	{}^{(A)}B^{i_1 i_2 \ldots i_k}_{\rm GR} = \frac{P(d_A|H_{i_1 i_2 \ldots i_k},\info)}{P(d_A|\hyp_{\rm GR},\info)},
\end{equation}
with $d_A$ being the data associated to the $A$th detection.

\subsection{Noise}
\label{subsec:noise}
From a theoretical point of view, the data favours the hypothesis $\hyp_{\rm modGR}$ compared to the hypothesis $\hyp_{\rm GR}$ when $\mathcal{O}^{\rm modGR}_{\rm GR} > 1$. The relative degree of belief in the two hypotheses is encapsulated in the magnitude of the odds ratio. However, in the case of advanced ground-based detectors, the signals will be buried deep inside the noise. This introduces the problem that the noise itself can mimic the effect of a deviation from GR that is non-negligible. 

Hence we need to study the effect of noise on the odds ratio. For this 
purpose, we constructed a so-called \textit{background}, \emph{i.e.} a distribution 
of log odds ratios from a large number of catalogues, collectively denoted by $\kappa$, of 
simulated signals consistent with $\mathcal{H}_{\mathrm{GR}}$ and embedded within
noise. The background distribution $P(\ln \mathcal{O}^{\rm modGR}_{\rm GR}|\kappa, \hyp_{\rm GR}, \info)$ of 
log odds ratios for catalogues of GR sources can be seen as the blue dotted histogram in the right hand panel of 
Fig.~\ref{fig:dphiA2_histograms} below.

In the advanced detector era, one will only have access to a single catalogue 
of detected \emph{foreground} events. The associated measured log odds ratio should 
subsequently be compared with the background distribution in order to quantify our
belief in a deviation from GR. To do this, in Sec.~\ref{subsec:catalogue_size} 
we will introduce a maximum tolerable false alarm probability $\beta$, which together 
with the background distribution sets a threshold $\ln \mathcal{O}_\beta$ for the
measured log odds ratio to overcome.

For specific examples of GR violations, we will want to know how likely it is that
the catalogue of foreground sources will have an odds ratio that is above threshold.
For this reason we will also simulate large numbers of foreground catalogues, collectively
denoted by $\kappa'$. For a given false alarm rate $\beta$, one can then calculate what 
fraction of the simulated foreground catalogues has a log odds ratio above the associated
threshold $\ln\mathcal{O}_\beta$; this fraction we will call the \emph{efficiency}.

\subsection{Implementation}
\label{subsec:implementation}
A few remarks have to be made regarding the implementation of the aforementioned method. First, we use $N_T = 3$ testing parameters $\{\psi_1, \psi_2, \psi_3\}$. The varying of these phase coefficients was parameterised in the following fashion:
\begin{equation}
	\psi_i = \psi_i^{\rm GR}(\mathcal{M},\eta)\,\left[1 + \delta\chi_i\right],
	\label{eq:deltapsi}
\end{equation}
with $\psi_i^{\rm GR}(\Mc,\eta)$ the functional form of the dependence of $\psi_i$ on $(\Mc,\eta)$ according to GR, and the dimensionless $\delta \chi_i$ is a fractional shift in $\psi_i$. The 0.5PN case, $\psi_1$, however cannot be implemented in a similar way, as GR predicts $\psi^{\rm GR}_1=0$. Instead, deviations from $\psi^{\rm GR}_1$ are modelled as
\begin{eqnarray}
	\psi_1^{\rm GR}(\Mc,\eta) = 0 &\rightarrow& \frac{3}{128 \eta} (\pi \Mc\,\eta^{-3/5})^{-4/3} \delta\chi_1,
	\label{eq:deltaps1}
\end{eqnarray}
and the interpretation of a fractional shift is not adequate; rather, $\delta\chi_1$ is related to the magnitude of the deviation itself.

For the computation of the odds ratio defined in Eq.~\eqref{eq:oddsindividual} and Eq.  \eqref{eq:oddscombined}, one needs to compute the relevant Bayes factors via the evidences. In high-dimensional problems, brute force methods to calculate the integral in Eq. \eqref{eq:evidence} are computationally too expensive. One can, however, make use of more efficient methods to make this calculation computationally feasible. In this paper, we resort to an algorithm called Nested Sampling \cite{Skilling:2004fu}. More specifically, an implementation tailored to ground-based observations of coalescing binaries by Veitch and Vecchio \cite{Veitch:2008fk, Veitch:2008qo, Veitch:2010tw} was used.

Both the model waveforms and the Nested Sampling algorithm were appropriately adapted from existing code in the LIGO Algorithms Library \cite{LAL}.


\section{Results}
\label{sec:results}
In this section, we want to lend further support to the claim in \cite{Li:2011fk} that the method is in principle sensitive to deviations that are not considered within the model waveforms, as long as the phase shift at $f \sim 150$ Hz, where the detectors are the most sensitive, is comparable to, say, a shift of $\delta\chi_3 \sim (\mbox{a few}) \times 10^{-2}$ at 1.5PN order (corresponding to a shift in the overall phase of $\sim 5$ radians at the given frequency). To this end we use two heuristic examples where the change in phase of the signals cannot be accomodated by the model waveforms, yet the deviation from GR turns out to be detectable.

The first example, in subsection \ref{subsec:mass_dep_freq_power}, considers an additional term in the phase associated with a power of frequency which \emph{itself} depends on the total mass of the system. This power is chosen in such a way that within the range of total masses we consider, the frequency dependence of the anomalous contribution varies from effectively being 0.5PN at the lower end to 1.5PN at the higher end. Clearly, our model waveforms are in no way designed to capture such a deviation from GR. The second case, in subsection \ref{subsec:quadratic_curvature}, considers a deviation at a PN order (2PN) that is higher than the orders at which we allow phase coefficients to vary in our model waveforms (0.5PN, 1PN, and 1.5PN). 

After presenting the main results, we study the effects of the number of detected sources on our confidence in a deviation from GR. For this investigation we use the example in subsection \ref{subsec:quadratic_curvature}.

From Fisher matrix analyses, it has been shown that the phase coefficients in Eq.~\eqref{psiidef} are best measured as the total mass of the system goes down 
\cite{Arun:2006ys, Arun:2006dz, Mishra:2010bs}. Therefore, the signals were 
chosen to originate from neutron stars with masses between $1\,\Ms$ and $2\,\Ms$. For such systems, it has been shown that contributions from the spin interactions and the sub-dominant signal harmonics are negligible, and merger/ringdown do not have a significant impact \cite{Van-Den-Broeck:2007zh,Van-Den-Broeck:2007zv, Vitale:2010cr}.

The aim is to simulate the situation at Advanced Virgo and LIGO as realistically as possible. We have assumed an advanced detector network with detectors at Hanford and Livingston, both with the Advanced LIGO noise curve \cite{Shoemaker:2009ct}, and a detector at Cascina with the Advanced Virgo noise curve \cite{Virgo-Collaboration:2009ys}. Three data streams were produced, containing stationary, Gaussian noise coloured by these respective noise curves, to which simulated signals were added. Events were placed uniformly in volume (\textit{i.e.} probability density proportional to $D^2$, where $D$ is the luminosity distance), between $100$ Mpc and $400$ Mpc, to reflect the estimates of the number of detectable sources and the appropriate horizon distance. A lower cut-off of 8 was 
imposed on the \emph{network} SNR, defined as the quadrature sum of the 
individual detector SNRs, so as to be consistent with the LIGO/Virgo minimum 
for an event to be claimed as detected.

The waveforms were chosen to go up to 2PN in phase both for the injected signals and the model waveforms. The test coefficients were taken to be $\psi_1$, $\psi_2$ and $\psi_3$, so that the hypothesis $\hyp_{\rm modGR}$ contains $2^{N_T}-1=7$ logically disjoint sub-hypotheses.

The priors given to the deviations $\delta\chi_i$ were chosen to be flat and centered around zero, with a total width of $0.5$. The priors on the remaining parameters were taken to be the same as in \cite{Veitch:2008fk}, with the exception that the distance is allowed to go up to $1000$ Mpc.

It should be stressed that the choice of waveform approximant, test coefficients, and priors on the deviations were, to a large extent, arbitrary. 
In the advanced detector era, one would seek to perform the most general test that computational resources will allow. This will include the most accurate waveforms available at that time, the highest number of test coefficients one can handle, and the least restrictive priors that are in accordance with our prior information at that moment.

\subsection{A deviation with a mass dependent power of frequency}
\label{subsec:mass_dep_freq_power}

In our first example, the signals are given a deviation in the phase that has a mass dependent frequency power. Specifically, the deviation is of the form:

\begin{equation}
	\Psi^{\rm GR}(\mathcal{M},\eta;f) \rightarrow \Psi^{\rm GR}(\mathcal{M},\eta;f) + \frac{3}{128 \eta} (\pi M f)^{-2+M/(3\Ms)},
	\label{eq:dchiA2}
\end{equation}
where $M$ denotes the total mass of the binary system. We note that for a system with component masses in the middle of our range, $(1.5,1.5)\,M_\odot$, the change in phase at $f = 150$ Hz is about the same as for a 10\% shift in $\psi_3$. More precisely, for these masses the change in $\Psi(\mathcal{M}, \eta; 150\,\mbox{Hz})$ is 13.3 radians, to be compared with the 12.8 radians change induced by a constant 10\% shift in $\psi_3$.

\begin{figure}[htbp!]
	\centering
	\includegraphics[angle=0,width=0.5\columnwidth]{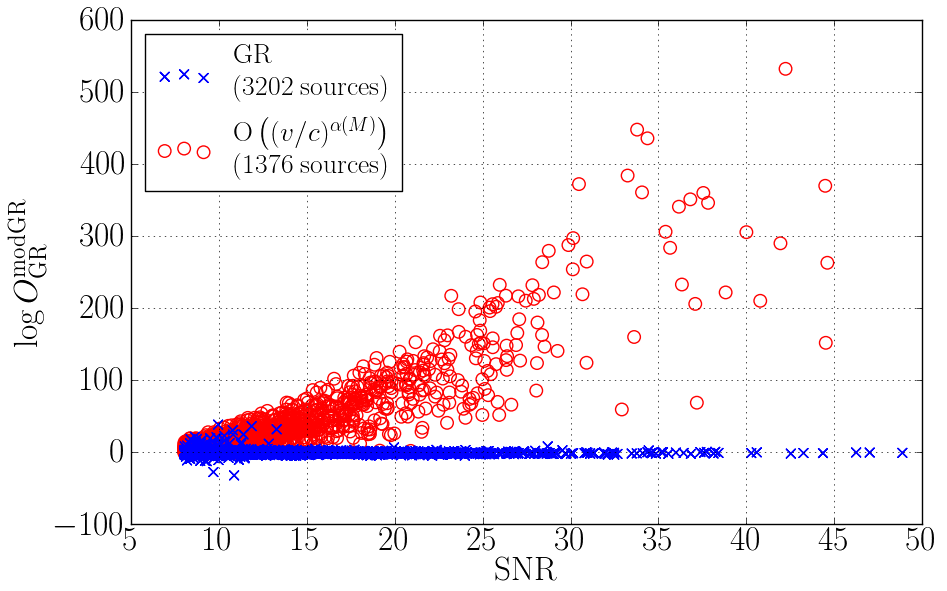}
	\caption{Distribution of log odds ratios as a function of the optimal network SNR. The crosses are for sources whose GW emission is in accordance with GR, while the circles are for sources with a deviation as in Eq.~\eqref{eq:dchiA2}. For the GR sources, the log odds ratios are concentrated tightly around zero, while for sources with the deviation in phase, they increase as a function of SNR. This is in agreement with the expectation that parameter estimation improves as the SNR increases. As our parameter estimation becomes better, deviations become more pronounced, which increases our confidence in a deviation from GR.}
	\label{fig:dphiA2_oddsVSsnr}
\end{figure}

In order to assess the statistics of the odds ratio, a large number or signals were simulated, with the parameter distribution explained above. For each of the signals, we calculated the odds ratio as defined in Eq. \eqref{eq:oddsindividual}. The distribution of the odds ratio as a function of SNR can be seen in Fig. \ref{fig:dphiA2_oddsVSsnr}. The separation between `foreground' and `background' is more or less complete already below SNR $\sim 15$. 

\begin{figure}[htbp!]
	\centering
	\includegraphics[angle=0,width=0.45\columnwidth]{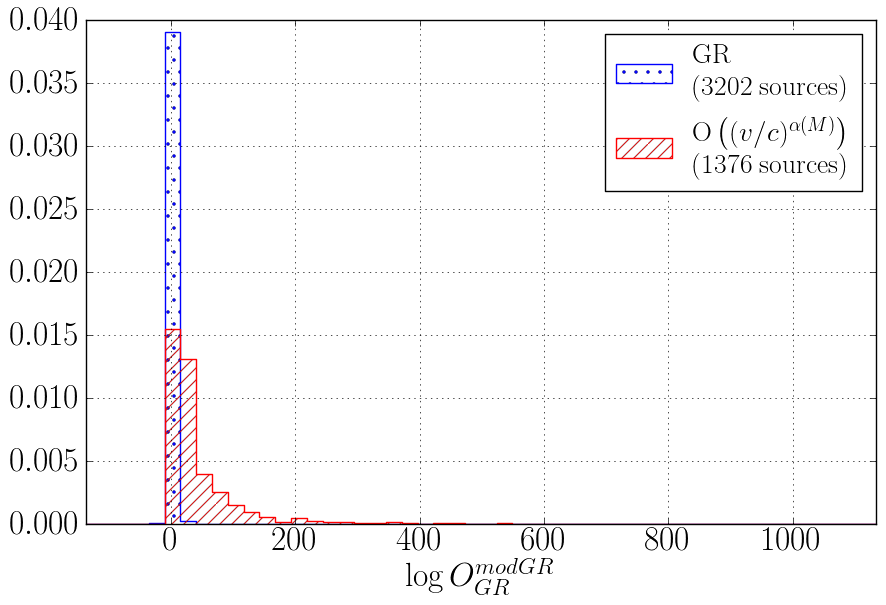}
	\includegraphics[angle=0,width=0.45\columnwidth]{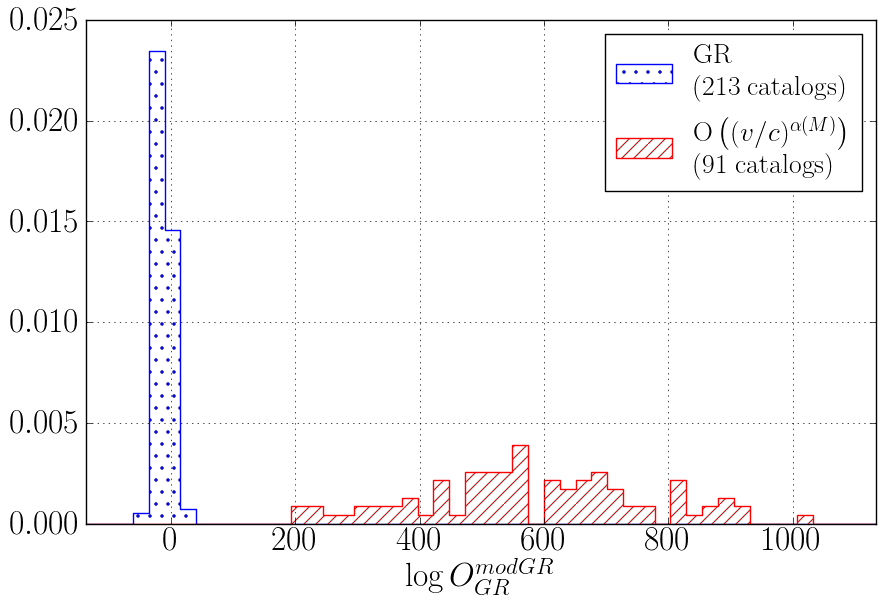}
	\caption{Left: Normalised distributions of log odds ratios for individual sources, where the signals are in accordance with $\hyp_{\rm GR}$ (blue dotted) or have a shift with a mass dependent frequency behaviour given in Eq.~\eqref{eq:dchiA2} (red striped). Right: Normalised distributions of logs of the combined odds ratios for the same injections as at the top, but collected into independent catalogues of 15 sources each. The effect of combining sources is to separate the distribution of GR injections and anomalous injections, increasing one's confidence in a deviation from GR.}
	\label{fig:dphiA2_histograms}
\end{figure}

In the top panel of Fig.~\ref{fig:dphiA2_histograms}, the odds ratios for the sources with a deviation from GR are compared with a `background' distribution (in the sense defined above). Next we collected sources into `catalogues' of 15 sources each and computed the \emph{combined} odds ratio of Eq.~(\ref{eq:oddscombined}) for all of these catalogues; the distribution of these odds ratios for `background' and `foreground' are shown in the bottom panel of Fig.~\ref{fig:dphiA2_histograms}. Clearly, the ability to combine information from multiple sources is a powerful tool in increasing one's confidence in a violation of GR. For the given deviation, a violation of GR can be established with near-certainty.

\subsection{A deviation at a higher PN order than the testing coefficients}
\label{subsec:quadratic_curvature}
Our testing coefficients are $\psi_1$, $\psi_2$, $\psi_3$, so that the model waveforms can only have shifts in PN phase contributions up to 1.5PN order. To show that we can nevertheless be sensitive to anomalies at higher PN order, we now consider signals with a constant shift at 2PN. We note in passing that theories with quadratic curvature terms in the action tend to introduce extra contributions at 2PN \cite{Yunes:2011uq, Stein:2010fk, Yagi:2011uq}.

Thus, we consider injections with 
\begin{equation}
	\psi_4 = \psi_4^{\rm GR}(\mathcal{M},\eta)\,\left[1 + \delta\chi_4\right],
	\label{eq:quadratic_curvature}
\end{equation}
where the magnitude is set to be $\delta\chi_4=0.2$. For comparison, at $f = 150$ Hz and for a system with component masses $(1.5, 1.5)\,M_\odot$, the change in the phase caused by such a deviation is comparable to the one caused by a negative relative shift in $\psi_3$ of 3.5\% (namely, the shift in $\Psi(\mathcal{M}, \eta, f)$ at $f = 150$ Hz is $\sim 4.5$ radians).

\begin{figure}[htbp!]
	\centering
	\includegraphics[angle=0,width=0.5\columnwidth]{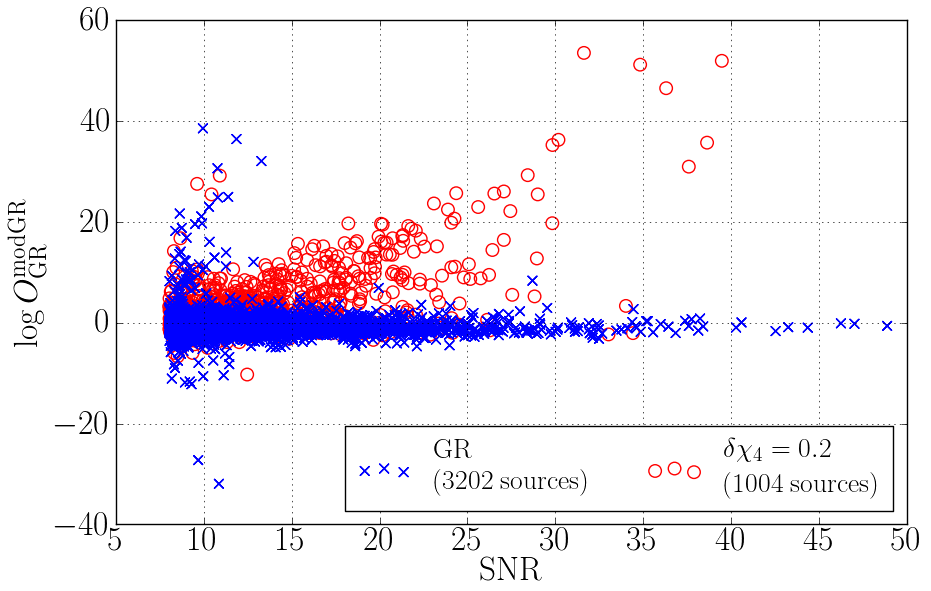}
	\caption{The distribution of odds ratio as a function of the optimal network SNR for GR injections (crosses) and injections with deviations from GR as in Eq.~\eqref{eq:quadratic_curvature} (circles). For the anomalous injections, the odds ratio increases as a function of SNR.}
	\label{fig:dphi4_20pc_oddsVSsnr}
\end{figure}

Fig.~\ref{fig:dphi4_20pc_oddsVSsnr} shows the odds ratio as a function of the optimal SNR, both for GR injections and anomalous ones. This time, as opposed to the example considered in subsection \ref{subsec:mass_dep_freq_power}, separation between the odds ratios of signals with the deformations of Eq.~\eqref{eq:quadratic_curvature} and the noise induced distribution of odds ratios for GR injections becomes apparent at SNR $\sim 20$. This can be attributed the fact that the deviations are in more subdominant contributions to the phase compared to the case considered earlier.

\begin{figure}[htbp!]
	\centering
	\includegraphics[angle=0,width=0.45\columnwidth]{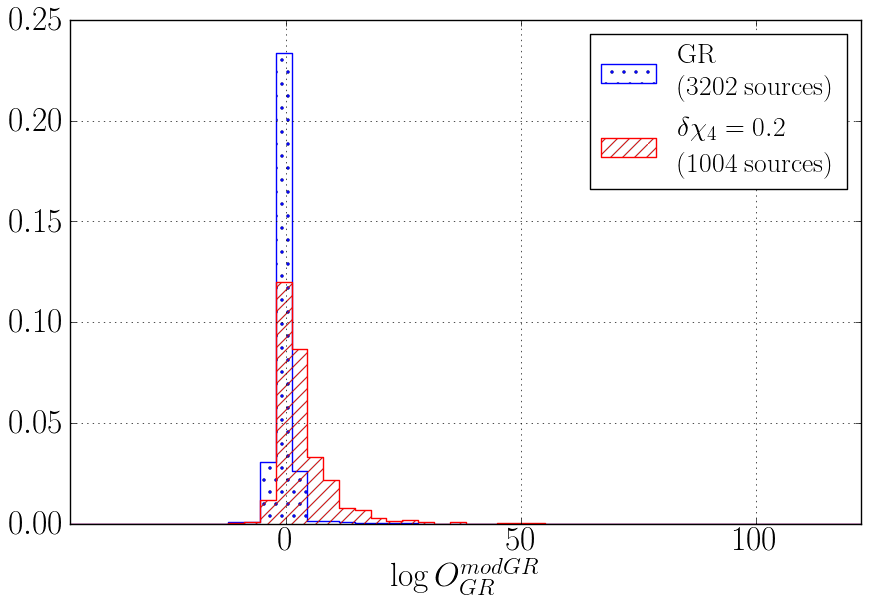}
	\includegraphics[angle=0,width=0.45\columnwidth]{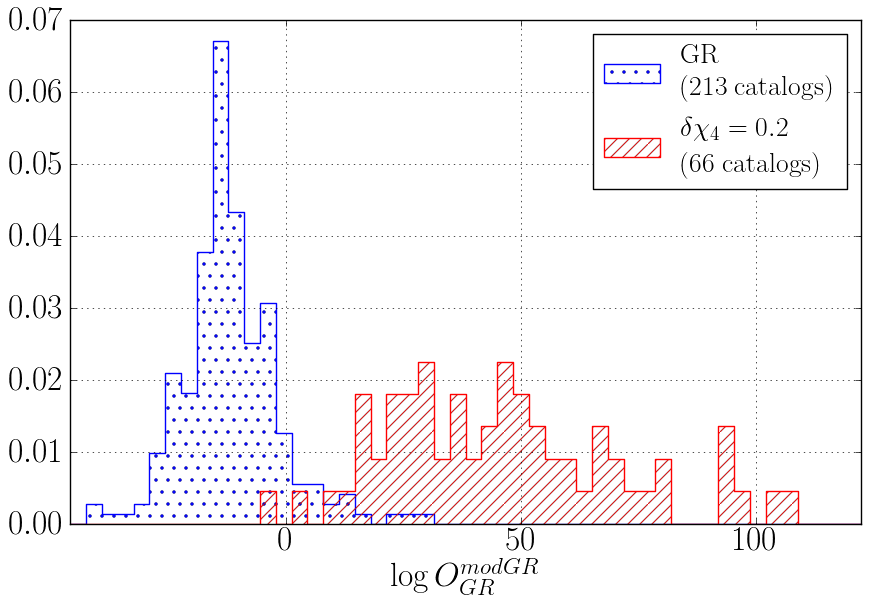}
	\caption{Left: normalized distribution of log odds ratios for individual sources, where the signals are in accordance with $\hyp_{\rm GR}$ (blue dotted) or have a deviation of the form given in Eq.~\eqref{eq:quadratic_curvature} (red striped). Right: normalised distribution of logs of the combined odds ratios for the same signals as at the top, but randomly arranged in catalogues of 15 sources each. The effect of combining sources is in this case is profound. Only a small difference between background and foreground is visible when considering individual sources. For catalogues of 15 sources, the differentiation becomes significant.}
	\label{fig:dphi4_20pc_histograms}
\end{figure}

In Fig.~\ref{fig:dphi4_20pc_histograms} we show the odds ratio for individual sources and for random catalogues with 15 sources each. For individual sources, the separation between the background and the foreground is present but weak. However, when one assumes random catalogues of 15 sources each, the separation becomes very significant. This further illustrates the importance of combining information from multiple sources.

\subsection{Effect of catalogue size}
\label{subsec:catalogue_size}
We have seen in the previous section that constructing the odds ratio from a catalogue of sources greatly increases the confidence in a deviation from GR. However, the rates of binary inspiral observed in the advanced detector era are highly uncertain. It is therefore instructive to study the effect of the catalogue size on our confidence in detecting a deviation.

To characterise such a confidence, we introduce the concept of \textit{efficiency}. Assume one has two distributions of log odds ratios: The \emph{background distribution} of log odds ratio obtained when the simulated catalogues, collectively denoted by $\kappa$, are in agreement with $\hyp_{\mathrm{GR}}$, $P(\ln\mathcal{O}^{\rm modGR}_{\rm GR}|\kappa, \hyp_{\mathrm{GR}}, I)$, and the \emph{foreground distribution} obtained when the simulated catalogues $\kappa'$ adhere to some alternative theory, $P(\ln\mathcal{O}^{\rm modGR}_{\rm GR}|\kappa', \hyp_{\mathrm{alt}}, I)$. Now choose a maximum tolerable \emph{false alarm probability} $\beta$. This sets a threshold $\ln\mathcal{O}_\beta$ for the measured log odds ratio to overcome, as follows:
\begin{equation}
	 \beta = \int_{\ln\mathcal{O}_{\beta}}^\infty P(\ln\mathcal{O}|\kappa, \hyp_{\mathrm{GR}}, \info)\, d\ln\mathcal{O}\,.
\end{equation}
We now define the \emph{efficiency}, $\zeta$, as the fraction of foreground with a false alarm probability of $\beta$ or less, \emph{i.e.} the portion that lies above the threshold $\ln\mathcal{O}_\beta$:
\begin{equation}
	\zeta = \int_{\ln\mathcal{O}_{\beta}}^\infty P(\ln\mathcal{O}|\kappa', \hyp_{\mathrm{alt}}, \info)\, d\ln\mathcal{O}.
\end{equation}
The efficiency can be viewed as the chance that if there is a deviation from GR corresponding to $\hyp_{\rm alt}$, the catalogue of sources that is actually detected will have a log odds ratio above threshold, \emph{i.e.} that it will have a false alarm probability of $\beta$ or less. 
Note that with these definitions, the efficiency is independent of the overall prior odds ratio $\alpha$ in Eqs.~(\ref{eq:totalignorance}) and (\ref{eq:oddscombined}), as it corresponds to the same shift of $\ln\alpha$ in the background distribution, the threshold $\ln\mathcal{O}_\beta$, and the foreground distribution.

\begin{figure}[htbp!]
\centering
\includegraphics[angle=0,width=0.5\columnwidth]{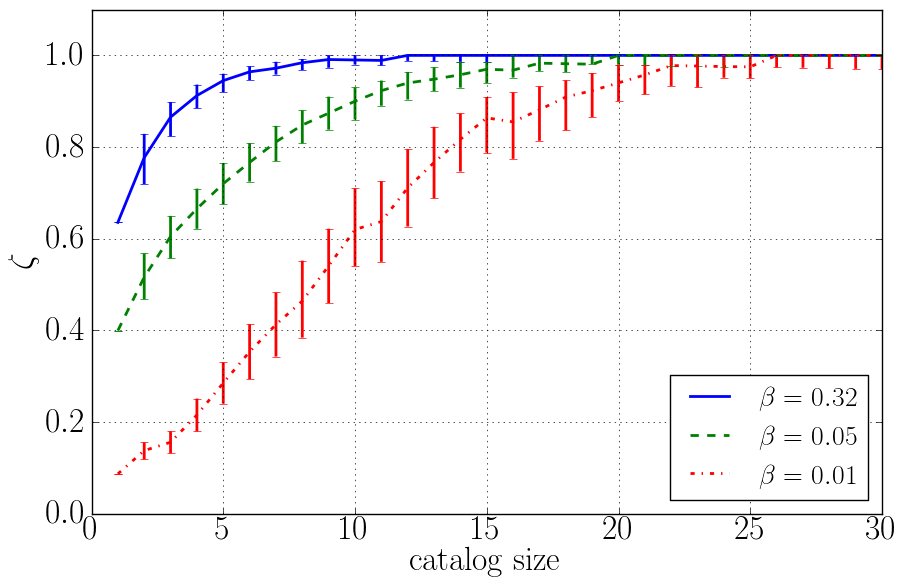}
\caption{The efficiency at a fixed false alarm rate as function of the catalogue size for the example described in subsection \ref{subsec:quadratic_curvature} ($\delta \chi_4 = 0.2$). 5000 random orderings of the same set of sources were split into catalogues. The mean (central curve) and the 68$\%$ confidence intervals (error bars) are plotted. The efficiency rises sharply as a function of the number of sources, underscoring the importance of coherently considering all detected signals events.}
\label{fig:sig_catsize}
\end{figure}

In Fig.~\ref{fig:sig_catsize}, we show the efficiency $\zeta$ for the example shown in subsection \ref{subsec:quadratic_curvature} as a function of the catalogue size, for $\beta \in \{0.32,0.05,0.01\}$. Which sources are placed together in a catalogue is determined randomly. To understand the statistical fluctuations in the efficiency when collecting sources into catalogues in different ways, for the same set of signals we considered 5000 random orderings in which the signals are combined into catalogues. The resulting median and the 68$\%$ confidence levels are shown as the central curve and the error bars, respectively. 

As can be seen in Fig.~\ref{fig:sig_catsize}, the acceptance probability rises sharply as a function of the catalogue size. This underscores the importance of considering all the detected source in a coherent fashion, as was explained in subsection \ref{subsec:multiple_sources}. Even though a single detection might not yield confidence in a deviation from GR, coherently adding information from multiple sources can rapidly increase this confidence.

To put the numbers in Fig.~\ref{fig:sig_catsize} into perspective, the predicted rate for binary inspiral in the so-called `realistic' case is 40 per year \cite{Abadie:2010zr}.


\section{Conclusions and discussion}
\label{sec:conclusion_discussion}

We have given two striking examples to support the claim that our method proposed in \cite{Li:2011fk} can distinguish deviations that are not captured by the limited model waveforms, as long as the phase shift in the frequency range where the detectors are the most sensitive ($f \sim 150$ Hz) is comparable to one caused by a shift of at least $\delta\chi_3 \sim (\mbox{a few}) \times 10^{-2}$, \emph{i.e.}, $\sim 5$ radians.

In the first example, signals were studied that have a deviation in the phase with a mass dependent power of frequency, effectively ranging from 0.5PN to 1.5PN as the total mass is varied from the lowest to the highest value we consider. The magnitude of the effect was such that at $f = 150$ Hz, the change in phase ($\sim 13$ radians) was about the same as that induced by a constant relative shift $\delta\chi_3 = -0.1$. The odds ratio for individual sources already showed confidence that such deviations can be measured. When sources were combined into catalogues of 15 each, the confidence in having detected a deviation improved drastically, and a deviation of this kind will be measurable with a false alarm probability of essentially zero.

We further showed results for signals with a shift in the 2PN phase coefficient, $\psi_4$. Setting $\delta\chi_4 = 0.2$, the induced change in phase at $f = 150$ Hz is comparable to a constant shift $\delta\chi_3 = -0.035$ at 1.5PN, namely 4.5 radians. The choice of a modification at 2PN was inspired by corrections to the phase if one considers a modified Einstein-Hilbert action containing terms that are quadratic in the Riemann tensor, as calculated in \cite{Yunes:2011uq, Stein:2010fk, Yagi:2011uq}. As can be seen from Fig.~\ref{fig:sig_catsize}, the efficiency for a maximum false alarm probability of 1\% is essentially unity for catalogues comprising more than 25  sources.

Lastly, we investigated the effect of the catalogue size on our confidence in detecting a deviation. In general, this confidence rises sharply with the number of sources in the catalogue, underscoring the necessity to combine information from multiple sources in the advanced detector era.

Finally, we want to mention some necessary future developments. First and foremost, the most accurate waveforms will need to be incorporated in order to distinguish between genuine effects predicted by GR, and possible deviations from GR. Especially for systems consisting of two black holes, or a neutron star and a black hole, these waveforms will need to include dynamical spins, sub-dominant signal harmonics, residual eccentricity, a description of the merger and ringdown, \textit{etc}. The development of waveforms including these effects is ongoing \cite{Barausse:2011bd, Ajith:2007il, Ajith:2011zt, Santamaria:2010jl, Buonanno:1999cq, Buonanno:2000kh, Damour:2000rq, Damour:2001fc, Buonanno:2006ss, Sturani:2010qc, Sturani:2010mw}. Furthermore, the effects of realistic detector noise, instead of idealised stationary, Gaussian noise, need to be studied in detail.

Once the advanced detectors have reached their design sensitivities and a number of detections have been made (\emph{e.g.} using template-based searches \cite{Babak:2006}), a test of General Relativity using compact binary coalescences could go as follows. Starting from the best available GR waveforms, one introduces parameterised deformations in phase as well as in amplitude, leading to disjoint hypotheses $H_{i_1 i_2 \ldots i_k}$, the logical `or' of all of which is $\mathcal{H}_{\rm modGR}$. Next, many injections are performed of GR waveforms into real or realistic data and collected into `catalogues' to establish a background distribution for the log odds ratio $\ln\mathcal{O}^{\rm modGR}_{\rm GR}$, and a suitable threshold $\ln\mathcal{O}_\beta$ is set below which a deviation from GR will not be accepted. Then $\ln\mathcal{O}^{\rm modGR}_{\rm GR}$ is computed for the catalogue of sources that were actually detected. If this number is above threshold, a violation of GR is likely.

The number of testing parameters one can consider will be limited, mainly by 
the computational restrictions one will have in the advanced detector era. Our 
method is meant, first and foremost, to establish whether or not a violation of 
GR is plausible, \emph{of whichever kind} and \emph{not} mainly to pinpoint
the eventual alternative theory of gravity responsible for the GW signal, nor
to estimate the parameters of the alternative model,
as it is unlikely that low-SNR signals as those expected for the advanced stage
of LIGO/Virgo will enable a detection of a GR deviation \emph{and} an 
identification of its nature. 
However, once a deviation is found, a follow-up investigation can be performed
with our inference method in an attempt to find out its precise nature by trying 
different alternatives to GR, \emph{i.e.} using waveforms inspired by specific (families of)
alternative theories of gravity. A version of the so-called parameterised post-Einsteinian 
waveform family \cite{Yunes:2009fu, Cornish:2011fk} could be useful in this respect. 
In this
regard we recall that our framework is not tied to any particular waveform family.

The results of \cite{Li:2011fk}, and the further investigations presented here, motivate the construction of a full data analysis pipeline based on the method we have presented. Although much work remains to be done on the data analysis side, the advanced detectors will enable us to go well beyond the tests performed using the observed binary pulsars, and give us our very first empirical access to the genuinely strong-field dynamics of space-time.

\section*{Acknowledgements}

TGFL, WDP, SV, CVDB and MA are supported by the research programme of the Foundation for Fundamental Research on Matter (FOM), which is partially supported by the Netherlands Organisation for Scientific Research (NWO). JV's research was funded by the Science and Technology Facilities Council (STFC), UK, grant ST/J000345/1. KG, TS and AV are supported by Science and Technology Facilities Council (STFC), UK, grant ST/H002006/1. The work of RS is supported by the EGO Consortium through the VESF fellowship EGO-DIR-41-2010. 

It is a pleasure to thank N.~Cornish, B.R.~Iyer, B.S.~Sathyaprakash and N.~Yunes for their valuable comments and suggestions. The authors would also like to acknowledge the LIGO Data Grid clusters, without which the simulations could not have been performed. Specifically, these include the computing resources supported by National Science Foundation awards PHY-0923409 and PHY-0600953 to UW-Milwaukee. Also, we thank the Albert Einstein Institute in Hannover, supported by the Max-Planck-Gesellschaft, for use of the Atlas high-performance computing
cluster.

\section*{References}

\end{document}